# Using and Designing Platforms for In Vivo Educational Experiments

**Joseph Jay Williams**
HarvardX, Harvard University
Joseph_jay_williams@harvard.edu

**Korinn Ostrow**
Worcester Polytechnic Institute
ksostrow@wpi.edu

**Xiaolu Xiong**
Worcester Polytechnic Institute
xxiong@wpi.edu

**Elena Glassman**
MIT
eglassman@mit.edu

**Juho Kim**
MIT
juhokim@mit.edu

**Samuel G. Maldonado**
San Jose State University
Samuel.g.maldonado@gmail.com

**Na Li**
Harvard University
nali01@fas.harvard.edu

**Justin Reich**
HarvardX, Harvard University
Justin_reich@harvard.edu

**Neil Heffernan**
Worcester Polytechnic Intstitute
nth@wpi.edu

## Abstract
In contrast to typical laboratory experiments, the everyday use of online educational resources by large populations and the prevalence of software infrastructure for A/B testing leads us to consider how platforms can embed *in vivo* experiments that do not merely support research, but ensure practical improvements to their educational components. Examples are presented of randomized experimental comparisons conducted by subsets of the authors in three widely used online educational platforms – Khan Academy, edX, and ASSISTments. We suggest design principles for platform technology to support randomized experiments that lead to practical improvements – enabling *Iterative Improvement* and *Collaborative Work* – and explain the benefit of their implementation by WPI co-authors in the ASSISTments platform.

## ACM Classification Keywords
H.4 Information Systems Applications; H.5 Information interfaces and presentation; K.3.1 Computer Uses in Education; J.4 Social and Behavioral Sciences

**Introduction**
Although the concept of A/B testing is widely known [Kohavi et al, 2009], using A/B testing to improve the content and interaction dynamics of educational resources like lessons and exercises is a relatively recent development in MOOCs. edX only released A/B testing of educational content this year, while it is still under development for Coursera.

## EXAMPLES OF A/B EXPERIMENTS IN KHAN ACADEMY, EDX, AND ASSISTMENTS

Examples are presented of how to conduct experiments in the platforms Khan Academy, EdX, and ASSISTments. These have been actually implemented by one or more of the co-authors. These provide context about the platform as well as insight into what kinds of variables can be manipulated and how randomization is implemented.

We then use this cross-platform comparison to elucidate two design principles that we propose platforms should support, and explain the benefits of designing ASSISTments to achieve these. These are *Iterative Improvement* and *Collaborative Work*. *Iterative Improvement* means the educational content can be repeatedly improved using the results of experiments and affordances of the technology. *Collaborative Work* refers to whether there is support for remote, asynchronous collaboration from diverse people looking at same resource, sharing insights and providing feedback on what any individual produces.

*1. Khan Academy*
Khan Academy makes available a documented open source framework for copying and creating new versions of math exercises. Any exercise at www.khanacademy.org/exercisedashboard can be downloaded as a single HTML file and then modified – to change the statement of the problem, the hints provided. An A/B test can then be performed that randomly assigns students to any version of the exercise with a weighted probability.

One example of a study conducted in collaboration with Khan Academy involved adding reflective question prompts above mathematics exercises which prompted students to explain while solving problems. The text was "clickable" so that when students clicked on it, new questions would appear. The original HTML file for the interactive exercise was simply copied, and then additional text and HTML code added into the copy. Four different versions of the exercise (with different questions) were created. These four files were then provided to Khan Academy, which used a weighted function to randomly assign participants to each kind of exercise whenever they visited it.

*2. EdX*
www.EdX.org is a platform for Massive Open Online Courses (MOOCs). A/B Testing or Content experiments in EdX involves declaring groups of participants (which must all be equal sized) and then inserting a component that can assign participants in different groups to different components.

The experiment conducted presented three different versions of an introductory motivational video to a MOOC. The first was spoken by the instructor for the course and emphasized specific steps students could take to keep themselves engaged and successful. The second was spoken by the instructor and was very positive and encouraging about their likely success, but did not provide specifics. The third had the exact same script from the first video - with specific steps students could take – but was spoken by *another* instructor. This allowed for measuring the effect of these different videos on subsequent behavior in the course.

*3. ASSISTments*
Any problem set or problem available on the ASSISTments platform can be copied via a button click into any user's own folder of exercises. It can then be modified using a WYSIWYG interface – changing not only hint text, but also what kind of targeted feedback is given for incorrect answers, and what problems are provided in subsequent attempts.

Any problem set can be copied and revised to create multiple versions, which are then nested within a Choose Condition problem set that randomly assigns participants to either version with equal probability. Because problem sets can be nested within problem sets, A/B testing can take place at several levels. However, it is less straightforward to link user information from one point in time to determine assigned condition at another point.

The ease with which A/B testing can be implemented within ASSISTments is highlighted by a study that was performed in only three weeks, from conception to data analysis and paper submission. Ostrow & Heffernan [2014] established an experimental design in which tutoring feedback on errors in a problem was provided as a video or as text. The benefits of this feedback on the subsequent problems of the same kind were assessed.

## ITERATIVE IMPROVEMENT

One of the key strengths of online resources is that they can be constantly updated and changed. Yet many platforms do not leverage this capacity to its full potential, or make it easy for continual editing and improvement of existing resources. For example, a large proportion of MOOCs are only slightly modified once produced, yet resources like Wikipedia are repeatedly edited and improved by an impressive crowd. Figure 1 shows the extent to which each of the three platforms support the design goal of Iterative Improvement.

ASSISTments supports Iterative Improvement because it provides a large library of existing exercises, and any lesson or interactive exercise within ASSISTments (shared via URL or ID) can be copied and modified using a WYSIWYG editor from a browser.

One way in which this has supported successive improvements was having a group of teachers look at 50 exercises from the library and create new hints and feedback them, just using their web browser and with no programming experience. These are now being tested in an ongoing experiment, and if shown to be effective, will be directly incorporated simply by clicking a button.

## COLLABORATIVE WORK

As extensive practical and research development in the field of crowdsourcing has revealed, one of the most powerful features of the Internet is that by collapsing typical spatial, temporal, and social barriers and distances between people, it leverages large and

| | Iterative Improvement | | |
|---|---|---|---|
| | Khan Academy | EdX | ASSISTments |
| Library of Existing Reusable & Modifiable Components | Yes | - Not accessible prior to course launch without instructor provision.<br>- Reuse of materials with permission of original instructor (and possibly institution) | Yes |
| Ease of Copying & Modifying | - Must download & initiate open source exercise framework<br>- Must know HTML | - Must first be invited as staff of target course<br>- WYSIWYG editor in EdX Studio | - Online WYSIWYG editor, regularly used by teachers with no programming knowledge |
| Mechanism for Replacing Existing Content | - Must first request pull from platform team for code review<br>- Then a straightforward change to A/B test | - Must first be granted role as staff for course<br>- Then a straightforward change to A/B test | - Straightforward change to A/B test made by user<br>- Must then email platform for data collection |
| Availability of Participants & Data for Repeated Iterations | Yes<br>- Each component is in continual use by new participants | No<br>- Each component can be used once in MOOC offering; cycle delays until MOOC is offered again | Yes<br>- Each component is used by teachers who assign it to their classes |
| | Collaborative Development | | |
| | Khan Academy | EdX | ASSISTments |
| Availability of Authoring Functionality | - Open Source exercise framework downloaded locally and used with knowledge of HTML | - Must be invited as staff of target course by institution in Consortium or approved for edge.edx.org.<br>- Cloud-based WYSIWYG functionality. | - Instant cloud-based WYSIWYG authoring tool |
| Affordances for Sharing Modifications & Eliciting Feedback | - Share hyperlinks to real demos via Internet server, and request feedback through typical channels (email, phone, in person) | - Cloud-based sharing of hyperlinks to real demos provided to those added as course beta users; request typical feedback.<br>- Solicit direct duplication or modifications using WYSIWYG authoring tool from those added as course staff. | - Cloud-based sharing of hyperlinks to real demos provided to anyone.<br>- Solicit direct duplication or modifications using WYSIWYG authoring tool from anyone with ASSISTments account.<br>- "Public Shared Folder" for all ASSISTments users |
| Proposals for A/B Tests & Access to Data | - Through central platform team<br>- External collaborations uncommon, but competition was held in April | - Cooperation of instructor required for both access & implementation | - Via platform team; submission process funded by NSF grant |
| Freedom to Run A/B Tests with Personal User Group | No<br>- Only exercise authoring | Yes<br>- Direct user group to MOOC, or re-implement course on edge.edx.org | Yes<br>- All content copied to one's folder can then be assigned to all student users.<br>- Mturk application |

diverse minds interacting through technology to make substantive improvements. Figure 1 shows how the platforms bring such strengths to bear, by supporting collaborative work on randomized experiments.

ASSISTments is designed so any user to share an exercise with any other user - not just a live demo, but the actual ability to copy and then modify an exercise.

This has facilitated over 10 external collaborations using ASSISTments directly between WPI and external researchers, bringing together diverse expertise in the studies conducted and resources created. In a Graduate course, 16 students were also able to form into teams of 4 and work collaboratively in person and asynchronously on designing experiments. They used the platform to receive direct feedback in the form of actual edits from the course instructor, a teaching assistant, and an actual math teacher.

An NSF Cyberlearning grant was awarded to WPI to support external researchers from multiple disciplines in proposing experiments to be run on the platform. In the period October through December, 2 complete studies were conducted and 4 are in progress, without any face-to-face meetings.